\documentclass[12pt]{article}

\usepackage{graphics,epsf}
\usepackage{amsfonts}
\usepackage[latin1]{inputenc}
\usepackage[small,bf]{caption}
\usepackage{amsthm}
\usepackage{amssymb}
\usepackage{amsmath}
\usepackage{cite}
\usepackage{graphicx}
\usepackage{bm}
\usepackage{epsfig}
\usepackage[dutch,english]{babel}
\usepackage{layout}
\usepackage{bibliotoc}
\usepackage{pstricks}
\usepackage{psfrag}
\usepackage{comment}

\newcommand{\nit}{\noindent}

\newcommand{\np}{\newpage}
\newcommand{\dsp}{\displaystyle}
\newcommand{\vs}[1]{\vspace{#1 ex}}
\newcommand{\hs}[1]{\hspace{#1 em}}
\newcommand{\bflr}{\begin{flushright}}
\newcommand{\eflr}{\end{flushright}}
\newcommand{\bc}{\begin{center}}
\newcommand{\ec}{\end{center}}
\newcommand{\ben}{\begin{enumerate}}
\newcommand{\een}{\end{enumerate}}

\newcommand{\be}{\begin{eqnarray}}
\newcommand{\ee}{\end{eqnarray}}
\newcommand{\ba}{\begin{array}}
\newcommand{\ea}{\end{array}}

\newcommand{\dd}[2]{\frac{\partial{#1}}{\partial{#2}}}
\newcommand{\ag}{\alpha}
\newcommand{\bg}{\beta}
\newcommand{\gam}{\gamma}

\newcommand{\eps}{\epsilon}

\newcommand{\thg}{\theta}

\newcommand{\sg}{\sigma}

\newcommand{\Del}{\Delta}
\newcommand{\Fg}{\Phi}

\newcommand{\Lb}{\Lambda}

\newcommand{\bfp}{{\bf p}}

\newcommand{\bfx}{{\bf x}}

\newcommand{\bfG}{{\bf G}}

\newcommand{\bfI}{{\bf I}}
\newcommand{\bfJ}{{\bf J}}

\newcommand{\bfP}{{\bf P}}

\newcommand{\lh}{\left(}
\newcommand{\rh}{\right)}

\newcommand{\beq}{\begin{equation}}
\newcommand{\eeq}{\end{equation}}

\newcommand{\veck}{{\bf k}}

\newcommand{\vecp}{{\bf p}}
\newcommand{\vecq}{{\bf q}}

\newcommand{\vecx}{{\bf x}}

\begin{document}

\pagestyle{empty}
\begin{flushright}
NIKHEF/2008-006
\end{flushright}

\bc
{\Large {\bf The Role of Conformal Symmetry}} \\
\vs{2}

{\Large {\bf  in the Jackiw-Pi Model}} \\
\vs{5}

{\large M.O.\ de Kok} 
\vs{2}

Inst.\ Lorentz, University of Leiden 
\vs{1}

P.O.\ Box 9506 
\vs{1}

2300 RA Leiden NL
\vs{1}

e-mail: mdekok@lorentz.leidenuniv.nl
\vs{3}

{\large J.W.\ van Holten} 
\vs{2}

Nikhef
\vs{1}

P.O.\ Box 41882
\vs{1}

1009 DB Amsterdam NL 
\vs{1}

e-mail: v.holten@nikhef.nl
\vs{3}

\today
\ec
\vs{5}

{\footnotesize
\nit
{\bf Abstract}\\
The Jackiw-Pi model in $2+1$ dimensions is a non-relativistic conformal 
field theory of charged particles with point-like self-interaction. 
For specific values of the interaction strengths the classical theory
possesses vortex and multi-vortex solutions, which are all degenerate
in energy. We compute the full set of first-order perturbative quantum 
corrections. Only the coupling constant $g^2$ requires renormalization; 
the fields and electric charge $e$ are not renormalized. It is shown 
that in general the conformal symmetries are broken by an anomalous 
contribution to the conservation law, proportional to the $\bg$-function. 
However, the $\bg$-function vanishes upon restricting the coupling 
constants to values $g^2 = \pm e^2$, which includes the case in which
vortex solutions exist. Therefore the existence of vortices also 
guarantees the preservation of the conformal symmetries. 
}

\np
\pagestyle{plain}
\pagenumbering{arabic}


\section{Introduction}

In this article the role of the conformal symmetries in the Jackiw-Pi model 
is studied at quantum level in $1$-loop approximation. 
The Jackiw-Pi model, first described in \cite{JP1,JP2,JP3}, can be seen as 
an extension of the non-linear Schr\"odinger model with $U(1)$ Chern-Simons 
gauge fields. A similar study for the non-linear Schr\"odinger model was 
performed in \cite{DeKokVanHolten}.

Like the non-linear Schr\"odinger model, a variation on the Jackiw-Pi 
model has been used to study the physics of gasses of bosonic particles 
\cite{Barashenkov}. The Jackiw-Pi model has also been considered in the 
context of Aharo\-nov-Bohm scattering \cite{BergmanLozano}, which
essentially is the scattering of charged particles in the plane by a 
magnetic flux tube. In reference \cite{BergmanLozano} the renormalization 
of coupling constant $g^2$ is studied. In this article the full 
renormalization at $1$-loop level is presented, and at this order of 
perturbation theory only $g^2$ is seen to renormalize: the matter and 
gauge fields, mass and electric charge $e$ are found not to renormalize.
 
The Jackiw-Pi model is a variation on the Abelian Higgs-model, with 
the Maxwell kinetic term replaced by a Chern-Simons term and 
with the matter Lagrangian taken to be non-relativ\-is\-tic. In the 
Abelian Higgs-model vortices arise as topological defects. 
For a specific choice of the coupling constants $e$ and $g^2$, i.e.\ 
$g^2 = -e^2$, stationary self-dual vortices arise in the Jackiw-Pi model 
as well \cite{JP1, JP2, JP3, Dunne, HorvCSLect}.
 
Like the non-linear Schr\"odinger model, the classical Jackiw-Pi model 
possesses, next to translation, rotation, Galilei and gauge invariance, 
also scale invariance and special conformal invariance. Due to the 
$1$-loop renormalization effects the scale and special conformal symmetries 
of the Jackiw-Pi model are seen to be broken precisely as in the non-linear 
Schr\"odinger model \cite{DeKokVanHolten}. However, this breaking does 
not occur when the renormalized coupling constants $e$ and $g^2$ are fixed 
to the value $g^2 = \pm e^2$. In particular, in the case of the minus sign 
both the conformal symmetries and the vortex solutions are seen to 
survive quantization.

This paper is structured as follows. Sect.\ 2 contains a short introduction 
to the Jackiw-Pi model and its vortex solutions. Sect.\ 3 reviews the 
symmetries of the model. Sect.\ 4 describes the quantization, whilst the 
1-loop corrections are calculated in sect.\ 5. In sect.\ 6 the scale 
dependence of the coupling constants is computed, and sect.\ 7 contains 
a summary and our conclusions.

\section{Introduction to the Jackiw-Pi Model}
\label{IntroJP}

The Jackiw-Pi model combines a Chern-Simons kinetic term for $U(1)$ gauge 
fields with the action of the non-linear Schr\"odinger model for a complex 
matter field $\Psi$ in $2$-dimensional Euclidean space 
\cite{JP1, JP2, JP3, Dunne, HorvCSLect}. 
The full action is given by
\be
S = \int dt d^2\vecx\,\,\frac{\sigma}{2}\epsilon^{\alpha\beta\gamma}
A_{\alpha}\partial_{\beta}A_{\gamma} + i\Psi^*D_t\Psi 
-\frac{1}{2}|{\bf D}\Psi|^2 - \frac{g^2}{2}(\Psi^*\Psi)^2,
\nonumber\\
\label{SJP}
\ee
where the mass $m$ of the matter field has been scaled away by rescaling 
the time variable by $t/m \rightarrow t$ \cite{DeKokVanHolten}. Furthermore, 
$D_{\alpha} = \partial_{\alpha}-ieA_{\alpha}$ is the gauge covariant 
derivative;  the indices in the Chern-Simons term run from $0$ to $2$ 
($\epsilon^{012}=1)$, $0$ denoting the temporal component of the gauge 
field and $1$ and $2$ its spatial components. The constant $\sigma$ equals 
$+1$ or $-1$, controlling the relative sign of the gauge kinetic term with 
respect to the other terms.

The equations of motion of this model are given by
\be
\label{EOM1}
&-i\partial_t\Psi = \frac{1}{2}{\bf D}^2\Psi + eA_0\Psi - g^2 (\Psi^*\Psi)\Psi, 
\ee
\be
\label{EOM2}
&\sigma(\nabla_1 A_2 - \nabla_2 A_1) + e \Psi^*\Psi = 0, 
\ee
\be
\label{EOM3}
&\sigma(\partial_tA_i-\nabla_iA_0) - \frac{ie}{2}\,\epsilon_{ij}
(\Psi^*D_j\Psi - (D_j\Psi)^*\Psi) = 0.
\ee
The equation of motion (\ref{EOM1}) is a gauged Schr\"odinger equation for 
the matter field. When the usual definitions for the magnetic field $B$, 
electric field $E_i$, charge density $\rho$ and charge current $J_i$
in two spatial dimensions are used,
\be
B \equiv \nabla_1 A_2 - \nabla_2 A_1, \quad E_i \equiv  \partial_tA_i-\nabla_iA_0,
\ee
\be
\rho \equiv \Psi^*\Psi, \quad J_i \equiv \frac{1}{2i}(\Psi^*D_i\Psi 
 - (D_i\Psi)^*\Psi).
\ee
the other two equations of motion (\ref{EOM2}) and (\ref{EOM3}) can be 
rewritten as
\be
B = -\sigma e\rho, \quad  E_i = - \sigma e\epsilon_{ij}J_j.
\label{GCS+}
\ee
These equations relate the magnetic field $B$ to the matter or charge 
density $\rho$ via the so-called Gauss-Chern-Simons equation, and the 
electric field to the  gauged matter or charge current $J_i$. The magnetic 
field is proportional to the matter density, and the electric field is 
perpendicular to the current, both with the factor $-\sigma e$. 

Next to the equations of motion that follow from the action there is also 
the Bianchi identity for the electromagnetic field. 
Normally in four dimensions this identity leads to the two source-free 
Maxwell equations, however in $2+1$ dimensions they reduce to a single 
equation:
\be
\partial_t B = \epsilon^{ij}\nabla_i E_j,
\ee
which in light of the equations above is nothing but the current 
conservation law:
\be
\partial_t \rho + {\bf \nabla}\cdot {\bf J} = 0.
\label{currentcons}
\ee

\noindent
It is well-known that under appropriate conditions the field equations 
(\ref{EOM1})-(\ref{EOM3}) admit stationary vortex-type solutions
\cite{JP1, JP2, JP3, Dunne, HorvCSLect}, such that $\partial_t\rho =0$. 
These solutions have strictly zero energy, as we show below. Under these 
circumstances the conditions for the existence of solutions can be derived 
from a Bogomol'nyi-type of argument. 

Indeed, defining $D_{\pm} = (D_1 \pm i D_2)$, the hamiltonian of the 
Jackiw-Pi Model can be written in the form
\be
H  & = & \int d^2\vecx \,\, \frac{1}{2}|{\bf D}\Psi|^2 + 
 \frac{g^2}{2}(\Psi^*\Psi)^2     \\\nonumber
   & = & \int d^2\vecx \,\, \frac{1}{2}|D_{\pm}\Psi|^2 \pm 
  \frac{eB}{2}\Psi^*\Psi + \frac{g^2}{2}(\Psi^*\Psi)^2.
\ee
Using the equation of motion (\ref{EOM2}), the energy of any
classical solution is then given by
\be
E & = & \int d^2\vecx \,\, \frac{1}{2}|D_{\pm}\Psi|^2 + \frac{1}{2}
\left(g^2 \mp \sigma e^2 \right)(\Psi^*\Psi)^2.
\ee
Therefore the absolute minimum $E = 0$ of the energy is reached 
with $\Psi \neq 0$ when 
\be 
D_{\pm} \Psi = 0, \hs{2}
g^2 \mp \sg e^2 = 0.
\label{JP-relation2}
\ee
Under these restrictions, solving the equations (\ref{EOM1})-(\ref{EOM3}) 
reduces to solving the Liouville equation
\be
(\nabla_1^2+\nabla_2^2)\log\rho = \pm \sigma 2 e^2 \rho.
\label{Liouville}
\ee
Solutions exist for a specific choice of sign, depending on $\sg$: 
$\pm \sigma = -1$. As a consequence it follows that 
\be
g^2 = - e^2 < 0.
\label{jw.1}
\ee
This relation is physically sensible: it states that repulsive
electromagnetic interactions are balanced by attractive contact 
interactions.

Explicitly, the solutions are given by 
\be
\rho = \frac{4}{e^2}\frac{|\partial_zf|^2}{(1+|f|^2)^2},
\label{rhoexpression}
\ee
where $f$ is a function of $z = x+iy$ only, allowing isolated poles 
(a meromorphic function). It is known that physically interesting 
solutions are given by $f$ being equal to a ratio of polynomials 
\cite{HorvCSLect, HorvathyYera}. These solutions show vortex behaviour
characterized by circulating currents $\bfJ$.


\section{The Symmetries of the Jackiw-Pi Model}

The Jackiw-Pi Model has a rich set of symmetries. Next to being gauge 
invariant, it is invariant under exactly the same group of space-time 
symmetries as the non-linear Schr\"odinger model, the Schr\"odinger group, 
\cite{DeKokVanHolten, Henkel, Niederer, JP1, JP2, JP3}. This group of 
transformations comprises space- and time-translations, rotations, Galilei 
transformations, scale transformations or dilatations and special conformal 
transformations.
The generators of Galilean boosts ${\bf G}$, conformal scaling $D$ and 
special conformal transformations $K$ are respectively given by
\be 
{\bf G} & =& t{\bf P} + \int d^2 \vecx \,\, \vecx |\Psi|^2;
\ee
\be
D = 2tH+ \int d^2\vecx \,\, x_i\left[\frac{i}{2}\Psi^*
\stackrel{\leftrightarrow}{\nabla}_i\Psi +\frac{\sigma}{2}
A_2\stackrel{\leftrightarrow}{\nabla}_iA_1  \right];
\ee
\be 
K = t^2H-tD-\frac{1}{2} \int d^2\vecx \,\, \vecx^2|\Psi|^2;
\ee
Noether's theorem then implies the conservation of the conformal
charges:
\be
\frac{dK}{dt} & =& -t\frac{dD}{dt} ; \quad \frac{dD}{dt} = 0.
\label{jw.2}
\ee
As in the case of the non-linear Schr\"{o}dinger model these
conservation laws imply relations between the field-dependent 
integrals 
\be
\ba{l}
\bfI \equiv \dsp{\int d^d x\, \bfx\, \Psi^* \Psi 
 = \bfG - t \bfP, }\\
 \\
I_1 \equiv \dsp{ \frac{1}{2}\, \int d^d x\, \bfx^2\, \Psi^* \Psi 
 = t^2 H - t D- K, }\\
 \\
I_2 \equiv \dsp{ \frac{i}{2}\, \int d^d x\, \bfx \cdot \Psi^* 
\stackrel{\leftrightarrow}{\bf D} \Psi = D - 2t H.}
\ea
\label{1.8}
\ee
Applying the conservation laws (\ref{jw.2}) and the conservation of
the hamiltonian, it follows that \cite{ghosh, DeKokVanHolten}
\be 
\frac{d\bfI}{dt} = - \bfP, \hs{2}
\frac{dI_1}{dt} = - I_2, \hs{2} \frac{dI_2}{dt} = - 2H.
\label{1.9}
\ee
As a result, the conformal invariance of the Jackiw-Pi model implies 
that any {\em stationary} solutions of the theory -- with $\rho$ 
time-independent, and for which $\bfI$, $I_1$ and $I_2$ are necessarily 
constant in time themselves -- necessarily have zero energy and momentum. 
This is the basis for the derivation of the classical solutions presented 
above.


\section{Quantization}
\label{JPrenorm}

In this section the Jackiw-Pi model is quantized using Feynman's path 
integral formulation. 

To be able to find a suitable propagator for the gauge fields, the gauge 
is fixed via the Faddeev-Popov method. Following \cite{BergmanLozano} the 
Coulomb gauge $\nabla_i A_i = 0$ is imposed by adding to the action a term
\be
\Del S_{GaugeFix} = \frac{1}{\xi}\int dtd^2\vecx\,\,\left(\nabla_iA_i\right)^2.
\ee
In the presence of this term the inverse of the kinetic term of the gauge 
fields in momentum space, following from action (\ref{SJP}) with the 
additional gauge fixing term, is given by
\be
\dsp{ 
\frac{\sigma}{\veck^4}\left(   
\begin{array}{ccc}
\xi k_0^2 & ik_2\veck^2 + \sigma\xi k_1k_0& -ik_1\veck^2 + \sigma\xi k_2k_0 
 \medskip\\\medskip
-ik_2\veck^2 + \sigma\xi k_1k_0 & \sigma\xi k_1^2 & \sigma\xi k_1k_2 \\
ik_1\veck^2 + \sigma\xi k_2k_0 &  \sigma\xi k_1k_2 & \sigma\xi k_2^2
\end{array} 
\right) }
\nonumber\\
\ee
Specializing to the Landau gauge $\xi =0$, the only nonvanishing component 
of the propagator of the gauge field describes propagation from $A_0$ to 
either $A_1$ or $A_2$ and vice versa. 

In \cite{BergmanLozano} it is explained that there are no real gauge 
particles in the Jackiw-Pi model since the gauge fields are completely 
constrained due to the equations of motion (\ref{EOM1})-(\ref{EOM3}). 
However, they can still be treated dynamically in internal lines. 
Another sign of this fact is that the propagator of the gauge field 
becomes independent of  $k_0$ in the Landau gauge, meaning that the 
propagator in coordinate space is instantaneous in time.

Fixing the gauge using the Faddeev-Popov method introduces additional 
Grassmannian fields to the action, the {\it ghost fields}. However in a 
$U(1)$ gauge theory these fields do not couple to the matter or gauge 
fields in Coulomb- or Lorentz-type gauges, and can consequently be ignored.
\\
\newline
Writing down the Feynman rules in momentum space in agreement with the 
conventions used in \cite{DeKokVanHolten}, it is seen that
each $\Psi$-propagator comes with
\be
\raisebox{-2.2ex}{\epsfig{file=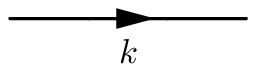, height=3.5ex}} =
\frac{1}{(2\pi)^{3}}\frac{i}{k_0 -\veck^2/2 +i\varepsilon};
\ee 
and each $A_{0}A_i$- or $A_iA_{0}$-propagator with 
\be
\raisebox{-2.2ex}{\epsfig{file=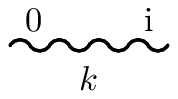, height=5ex}} =
-\,\raisebox{-2.2ex}{\epsfig{file=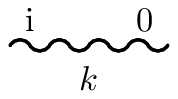, height=5ex}} =
\frac{\sigma}{(2\pi)^{3}}\frac{\epsilon^{ij}k_j}{ \veck^2}.
\label{A0iprop}
\ee 
In the Jackiw-Pi model there are two $3$-point vertices, one involving the 
$A_0$ field and one involving either $A_1$ or $A_2$. The Feynman rule 
for the first is given by
\be
\raisebox{-7ex}{\epsfig{file=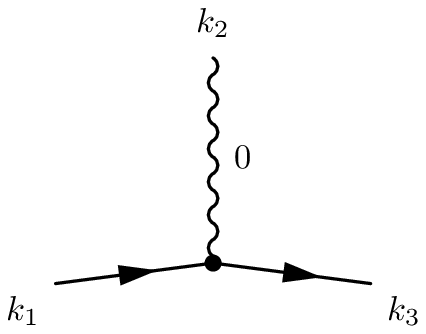, height=18ex}} = -ie (2\pi)^{d+1}, 
\ee
and for the latter by
\be
\raisebox{-7ex}{\epsfig{file=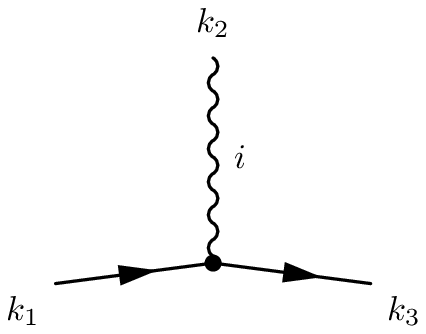, height=18ex}} = 
-i\frac{e}{2}(\veck_1+\veck_3)(2\pi)^{d+1}.
\ee
Also there are two $4$-point vertices. Next to the one also present in the 
non-linear Schr\"odinger model, 
\be
\raisebox{-8ex}{\epsfig{file=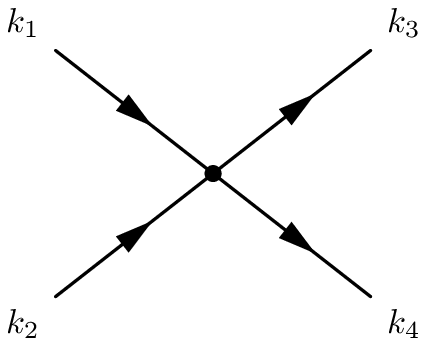, height=18ex}} = 
i\frac{g^2}{2}(2\pi)^{d+1},
\ee
one involving twice the gauge field $A_1$ or $A_2$:
\be
\raisebox{-8ex}{\epsfig{file=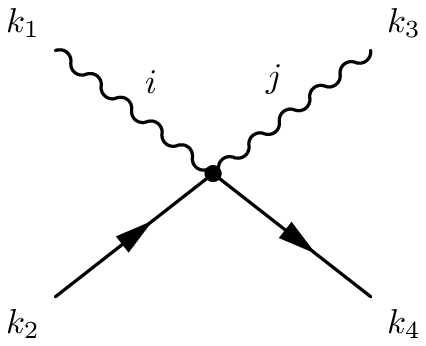, height=18ex}} = 
i\frac{e^2}{2}\delta^{ij}(2\pi)^{d+1}.
\ee
Finally, all vertices naturally come with energy and momentum conservation.

\section{Renormalization Effects}

Next the perturbative corrections to the tree-level propagators and 
coupling constants are calculated at the $1$-loop level. The procedure 
is similar to that used in \cite{DeKokVanHolten}: first the $k_0$-integral 
is performed, and then dimensional regularization is used to perform the 
remaining $\veck$-integral.

As it turns out, most Feynman diagrams vanish at $1$-loop level. This can 
be traced back to the fact that the propagator of the gauge field 
does not depend on $k_0$ and is an odd function of $\veck$, equation 
(\ref{A0iprop}).

\subsubsection*{Corrections to the $A$-propagator}

The only diagram that can be written down for the propagator of the gauge 
field at one loop level is given by
\be
\raisebox{-1.5ex}{\epsfig{file=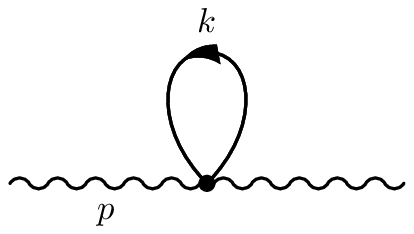, height=12ex}}
\ee
This {\it seagull diagram} is known to vanish, as was discussed in 
\cite{DeKokVanHolten}.
It is thus concluded that the gauge field is not renormalized.

\subsubsection*{Corrections to the $\Psi$-propagator}

Knowing that the seagull diagram vanishes, the only possible correction 
to the $\Psi$ propagator comes from the diagram
\be
\raisebox{-2ex}{\epsfig{file=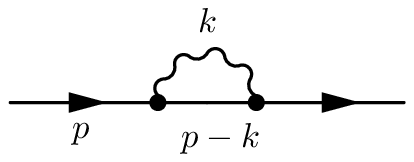, height=10ex}}
\ee
where the photon propagator either runs from $A_0$ to $A_i$ or vice versa. 
Since these contributions are equal up to a minus sign, the sum of these
two diagrams vanishes. However, on their own the diagrams are also seen to 
vanish. Indeed, each of the diagrams is proportional to
\be
\int dk_0d^2\veck\,\, \frac{2p_i-k_i}{p_0-k_0-\frac{1}{2}(\vecp-\veck)^2
+ i\varepsilon}\frac{\sigma \epsilon^{ij}k_j}{ \veck^2}.
\ee
After performing the $k_0$ integral, one is left with 
\be
\pi i\int d^2\veck\,\frac{\sigma\epsilon^{ij}(2p_i-k_i)k_j}{\veck^2} = 0,
\ee
which equals zero because $\epsilon^{ij}k_ik_j = 0$ and the oddness in 
$\veck$ of the remaining term.

It follows that the $\Psi$-field is not renormalized at this order 
in perturbation theory.

\subsubsection*{Corrections to the $A_0$-3-point vertex}

The only possible diagram is given by 
\be
\raisebox{-1.5ex}{\epsfig{file=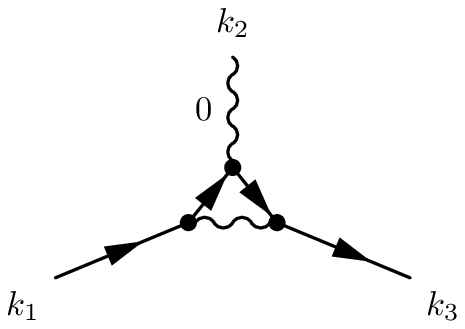, height=18ex}}
\ee
where the $k_i$ are the three-momenta $(k_{i,0}, \veck_i)$ running 
through the different external propagator lines.

Because the propagator of the gauge field doesn't depend upon $k_0$, 
the integral over $k_0$ is of the form
\be
\int dk_0 \frac{1}{k_0+\ldots-i\varepsilon}\frac{1}{k_0+\ldots-i\varepsilon} 
 = 0.
\ee
Performing the integral via closing of the integration contour in the 
complex plane, there are two poles, both in the upper half plane. 
Closing the contour in the lower half plane, the integral is directly 
seen to vanish. This means the whole diagram vanishes and that 
the coupling constant $e$ of the $A_0$-3-point vertex, the electric charge, 
receives no corrections. For consistency this means that also the other 
$3$-point vertex and the $4$-point vertex with the gauge fields should 
receive no corrections, since they are proportional to $e$ and $e^2$ 
respectively.

\subsubsection*{Corrections to the $A_i$-3-point vertex}

Indeed, also the corrections to the $3$-vertex involving $A_i$ are 
seen to vanish. The possible diagrams are given by 
\be
&\raisebox{-8ex}{\epsfig{file=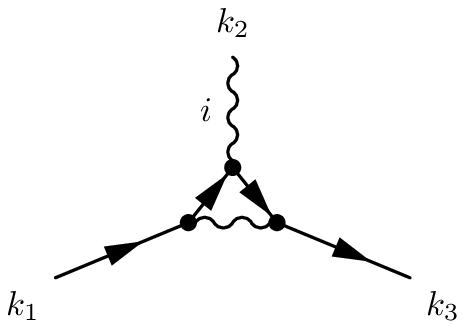, height=18ex}} + 
 \\\nonumber
&\raisebox{-8ex}{\epsfig{file=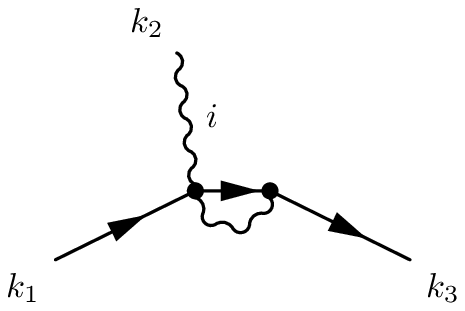, height=18ex}} +
\raisebox{-8ex}{\epsfig{file=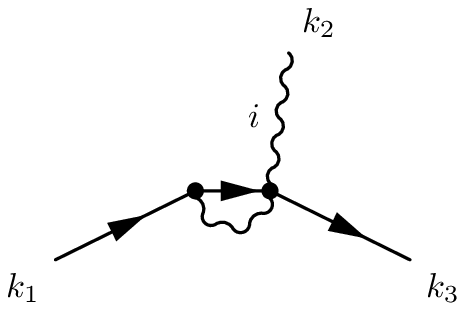, height=18ex}} 
\ee
The first vanishes due to the $k_0$ integral in the complex plane, the 
second and third because after the $k_0$ integral an integral odd in 
$\veck$ remains.

\subsubsection*{Corrections to the $A^2\Psi^2$-4-point vertex}

In this case the one loop level diagrams are given by
\be
&\raisebox{-8ex}{\epsfig{file=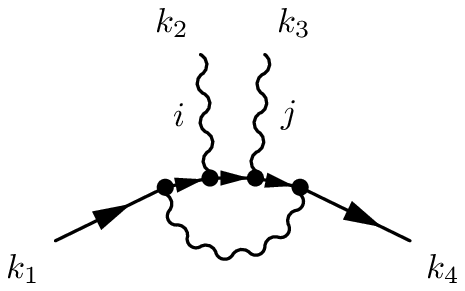, height=18ex}} + \\\nonumber
&\raisebox{-8ex}{\epsfig{file=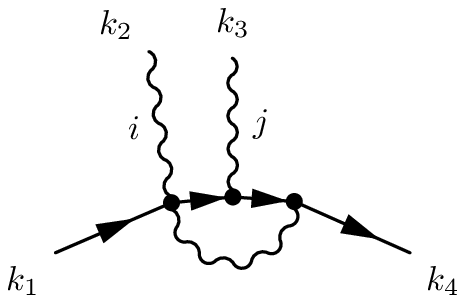, height=18ex}} + 
\raisebox{-8ex}{\epsfig{file=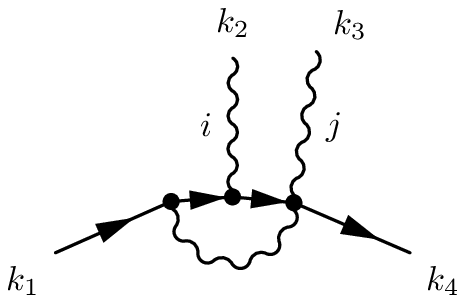, height=18ex}}
\ee
All three are seen to vanish because the photon propagator does not depend 
on $k_0$.

Thus the complete set of relevant vertices consistently implies that the 
electric charge $e$ is not renormalized at 1-loop level.

\subsubsection*{Corrections to the $\Psi^4$-vertex}

The only renormalization effects in the Jackiw-Pi model at 1-loop level 
appear when calculating the corrections to the coupling constant $g^2$ of 
the $\Psi^4$-interaction.

At tree level the contributions to this process are given by
\be
&\raisebox{-8ex}{\epsfig{file=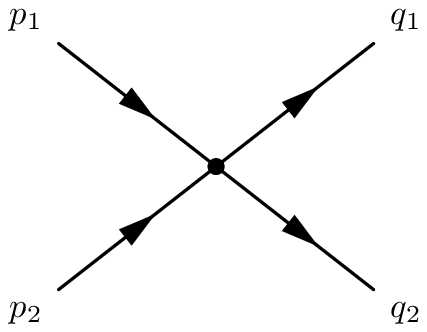, height=18ex}} +  \\\nonumber
&\raisebox{-8ex}{\epsfig{file=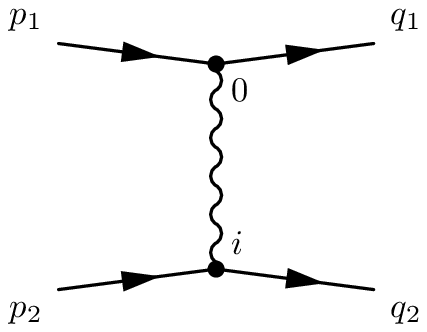, height=18ex}} +  
\raisebox{-8ex}{\epsfig{file=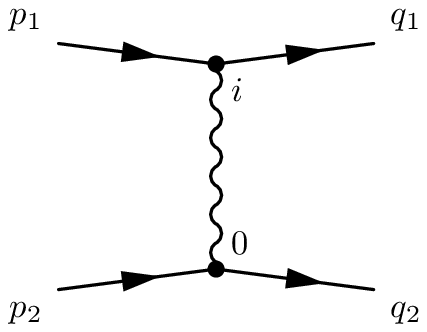, height=18ex}} 
\ee
The expression corresponding to this sum of diagrams has been calculated in 
\cite{BergmanLozano} working in the center of mass frame.
Up to the terms that describe the external lines,  it is in the present 
conventions given by
\be
A^0(p_i,q_j) =  i\frac{g^2}{2} -\frac{\sigma e^2}{2}\cot\theta,
\ee
where $\theta$ is here defined as the scattering angle between the colliding 
particles in the plane.

At one loop level the four contributing diagrams are given by 
\be
\raisebox{-8ex}{\epsfig{file=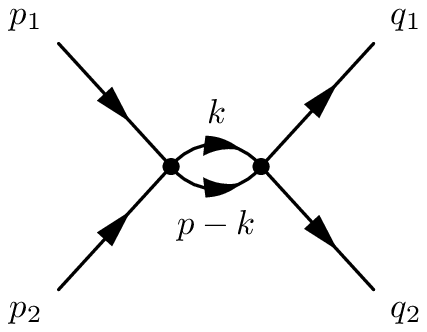, height=18ex}} +
\raisebox{-8ex}{\epsfig{file=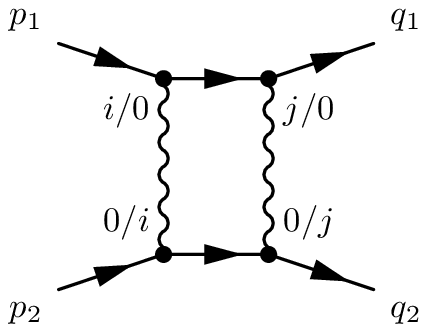, height=18ex}} +  \\[5pt]\nonumber
\raisebox{-8ex}{\epsfig{file=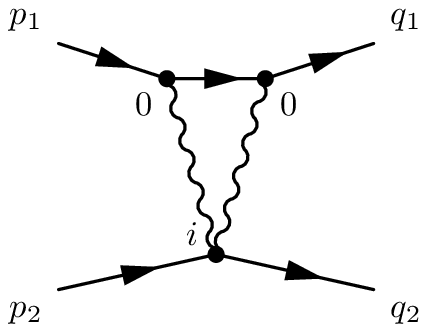, height=18ex}} +  
\raisebox{-8ex}{\epsfig{file=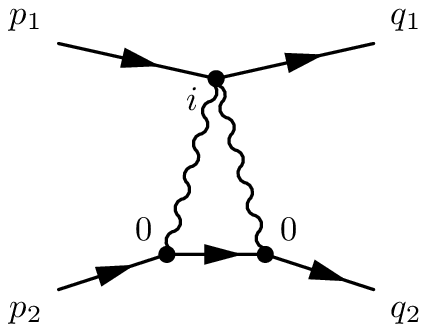, height=18ex}}  \phantom{+}
\ee
The upper right box diagram represents four different diagrams, depending 
on where the vertices involving $A_0$ are placed. The corresponding 
expressions have been examined in \cite{BergmanLozano} and found to be finite. 
The total result is given by
\be
A^{(1)}_{box} = \frac{i\Lb^{-\eps}}{8\pi m} e^4 
\left(\ln|2\sin\theta| - i\pi\right).
\ee
Here we have extracted the dimension of the amplitude in $d = 2 - \eps$
dimensions using an arbitrary momentum scale factor $\Lb$.
 
Also the two triangle diagrams have been studied in \cite{BergmanLozano}. 
In that paper their divergence is controlled via a cut-off in momentum space. 
Here they are studied using dimensional regularization as was done in chapter 
\cite{DeKokVanHolten}.

The bottom left triangle diagram reads
\be
&&A^{(1)}_{triangle,1} = (-ie)^2\left(\frac{ie^2}{2}\right) \int 
\frac{dk_0d^2\veck}{(2\pi)^3} B_{triangle,1} \\
&&B_{triangle,1} = \frac{\epsilon^{in}k_n}{\veck^2}
\frac{\epsilon^{i}(q_2-p_2-k)_m}{(\vecq_2-\vecp_2-\veck)^2} \times \\\nonumber
&&\quad\quad\quad\quad\quad\quad\quad
\frac{i}{p_{1,0}-k_0-\frac{1}{2}(\vecp_1-\veck)^2+i\varepsilon}
\ee
Performing the $k_0$ integral and writing the integrand into one denominator 
using Feynman's trick, leads to
\be
A^{(1)}_{triangle,1} = \frac{e^4i}{4}\int \frac{d^2\veck}{(2\pi)^2} \int_0^1dx\, 
\frac{\veck^2-x(1-x)(\vecp_2-\vecq_2)^2}{\left[
\veck^2+x(1-x)(\vecp_2-\vecq_2)^2 \right]^2}.
\ee
Using the technique of dimensional regularization, this integral becomes
\be
A^{(1)}_{triangle,1} = \frac{e^4i}{16\pi} 
\Lambda^{-\epsilon}\left\{\frac{2}{\epsilon}-\gamma_E -\frac{1}{4\pi} 
+2-\ln\left(\frac{(\vecp_2-\vecq_2)^2}{4\pi\Lambda^2}\right)  \right\} ,\nonumber\\
\label{dimregtriangle1}
\ee
where the momentum scale has been extracted. Similarly, the other triangle 
diagram leads to
\be
A^{(1)}_{triangle,2} = \frac{e^4i}{16\pi} \Lambda^{-\epsilon}
\left\{\frac{2}{\epsilon}-\gamma_E -\frac{1}{4\pi} + 
2-\ln \left(\frac{(\vecp_1-\vecq_1)^2}{4\pi\Lambda^2} \right) \right\} \nonumber\\
\label{dimregtriangle2}
\ee

In the Center of Mass frame where $\vecp_1 = -\vecp_2 = \vecp_{CM}$ and 
$\vecq_1 = -\vecq_2 = \vecq_{CM}$ the results (\ref{dimregtriangle1}) and 
(\ref{dimregtriangle2}) are equal and given by
\be
A^{(1)}_{triangle,CM} = \frac{e^4i}{16\pi} \Lambda^{-\epsilon}
\left\{\frac{2}{\epsilon}-\gamma_E -\frac{1}{4\pi} + 
2-\ln \left(\frac{\vecp^2_{CM}\sin^2\theta/2}{\pi\Lambda^2} \right) \right\},
\nonumber\\
\label{dimregtriangleCM}
\ee
where (as before) $\theta$ is the scattering angle in the plane.

Using the result obtained in \cite{DeKokVanHolten} for the remaining diagram, 
which in the Center of Mass frame and in the present conventions is given by
\be
A^{(1)}_{NLS} &=& -\frac{g^4i}{8\pi}\Lambda^{-\epsilon}\left\{\frac{2}{\epsilon}
-\gamma_E - i\pi 
- \ln\frac{\vecp^2_{CM}}{4\pi\Lambda^2} \right\}, \\\nonumber
\ee
the total result at one loop level is given by
\be
A^{(1)} &=&  \frac{ig^2}{2} -\frac{\sigma e^2}{2}\cot\theta 
+ \frac{i \Lb^{-\eps}}{8\pi} e^4
\ln|2\sin\theta| \\\nonumber
&&- \frac{i\Lambda^{-\epsilon}}{8\pi}\left[g^4- e^4 \right]
\left(\frac{2}{\epsilon}-\gamma_E - i\pi\right)
 \\\nonumber
&&+\frac{i\Lambda^{-\epsilon}}{8\pi}g^4\ln \frac{\vecp^2_{CM}}{4\pi\Lambda^2}  
-\frac{i\Lambda^{-\epsilon}}{8\pi}e^4\left[\frac{1}{4\pi} -
2 + \ln\frac{\vecp^2_{CM}\sin^2\theta/2}{\pi\Lambda^2}  \right].
\ee
This expression is the result of a subtle interplay between terms coming 
from the different diagrams. For example, the $2/\epsilon$ term
comes from the triangle diagrams and the $1$-loop diagram from the 
non-linear Schr\"odinger model, while the $i\pi$ term comes from this
latter diagram and the box diagram, yet both $2/\epsilon$ and $i\pi$ come 
with the same coefficient.

The $1$-loop result for $A$ clearly diverges as $\epsilon \rightarrow 0$. 
This divergence can be repaired by introducing renormalized coupling 
constants:
\be
g^2 \Lb^{-\eps} = g_R^2 + \frac{1}{4\pi} \lh \frac{2}{\eps} - \gam_E \rh
\lh g_R^4 - e_R^4 \rh, \hs{2} e^2 \Lb^{-\eps} = e_R^2,
\label{gRlambdaJP}
\ee
where $g_R^2$ and $e_R^2$ are dimensionless. 
In terms of this renormalized coupling constant $g_R^2$, $A^{(1)}$ is finite 
and given by
\be
\Lb^{-\eps} A^{(1)} &=& 
 \dsp{ \frac{ig_R^2}{2}-\frac{\sigma e_R^2}{2}\cot\theta }\\\nonumber
&& \dsp{ - \frac{ie_R^4}{8\pi}
 \left( \frac{1}{4\pi} -2 + \ln \left| \tan \frac{\thg}{2} \right| \right), }\\\nonumber
&& \dsp{ + \frac{i}{8\pi}\left( g_R^4- e_R^4 \right) \left[ i\pi 
 + \ln \frac{\bfp^2_{CM}}{4\pi \Lb^2} \right] }
\ee
in the limit $\epsilon \rightarrow 0$.

\section{The Running Couplings}

The 1-loop renormalization (\ref{gRlambdaJP}) leads to a scale dependence 
of the coupling constants given in the limit $\eps \rightarrow 0$ by the 
$\bg$-functions
\be
\bg_e (g_R^2, e_R^2) \equiv \Lambda \dd{e_R^2}{\Lb} = 0, \hs{2} 
\beta_g(g_R^2, e_R^2) \equiv \Lambda \frac{\partial g_R^2}{\partial \Lambda} 
 = \frac{1}{2\pi}\left(g_R^4-e_R^4 \right).
\label{6.1}
\ee
Solving these equations leads to a constant electromagnetic coupling 
$\ag = e_R^2$, and a running coupling constant $g_R^2(\Lb)$, except for the 
special values 
\be
g_R^2(\Lambda) \equiv \pm \ag,
\ee
one of which is precisely the condition necessary for the existence
of time-independent self-dual solutions (\ref{jw.1}). 

\begin{figure}[t]
\hfil
\includegraphics[width=80mm]{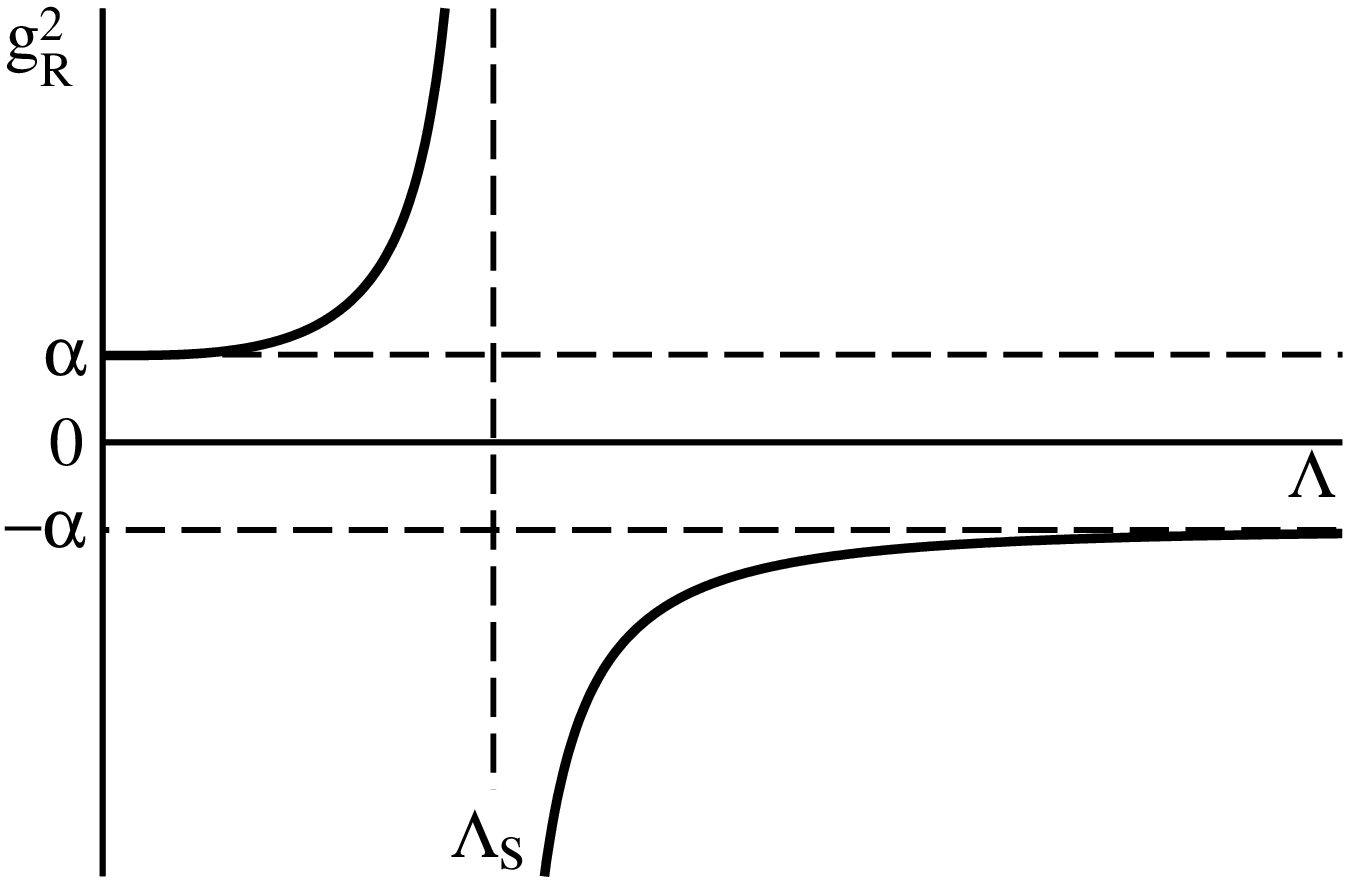}
\caption{Plot of $g_R^2$ as a function of the momentum scale
 $\Lambda$ for $|g_R^2(\Lambda_*)| > \alpha$.}
\label{grplot2a}
\end{figure}

\begin{figure}[t]
\hfil
\includegraphics[width=80mm]{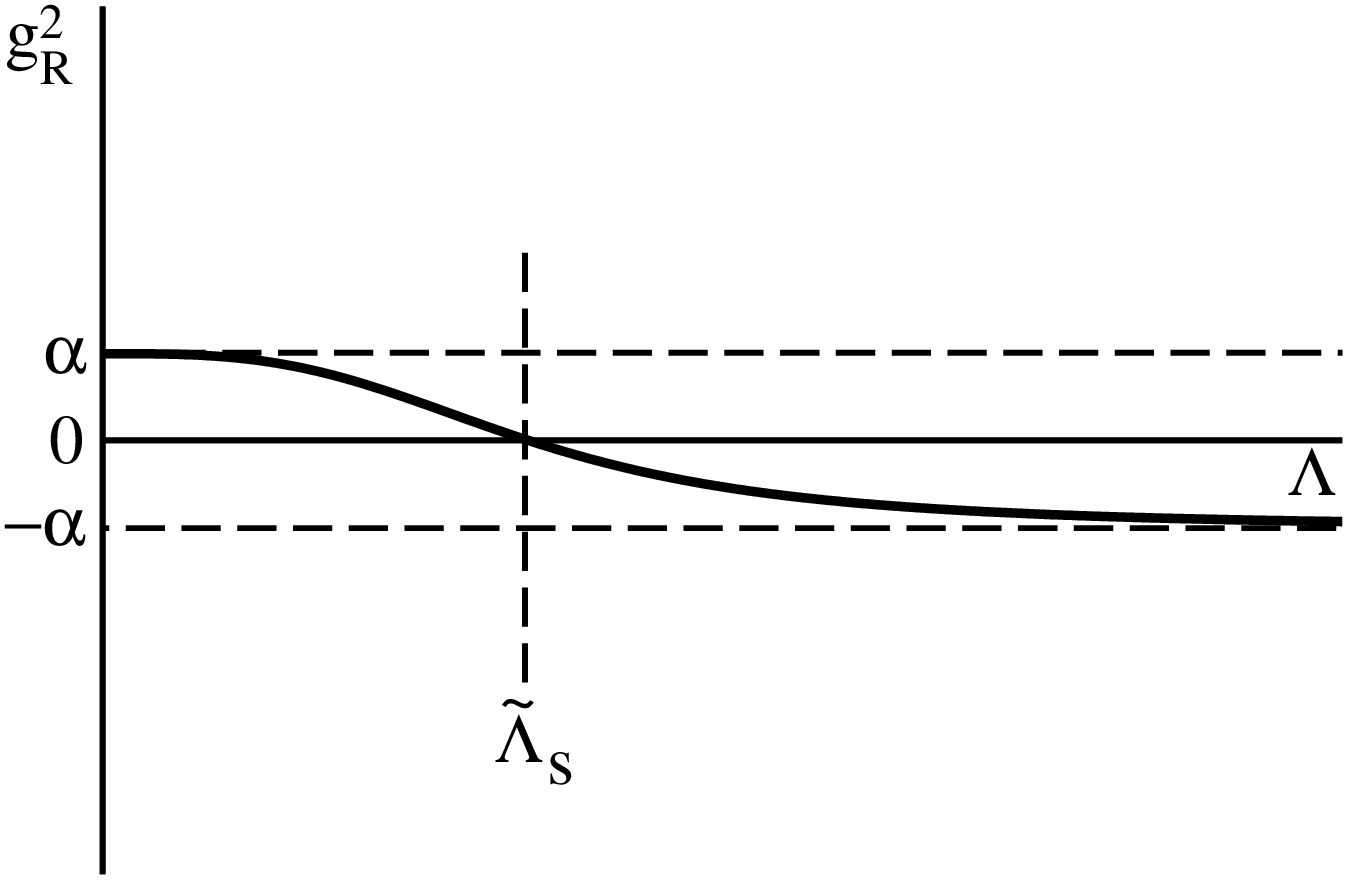}
\caption{Plot of $g_R^2$ as a function of the momentum scale $\Lambda$ 
 for $|g_R^2(\Lambda_*)| < \alpha$}
\label{grplot2b}
\end{figure}

Whenever $g_R^2(\Lb) \neq \pm \ag$, the solution of equation (\ref{6.1}) 
is given by
\be
\left| \frac{g_R^2(\Lb) - \ag}{g_R^2(\Lb) + \ag} \right| = 
\left| \frac{g_*^2 - \ag}{g_*^2 + \ag} \right| 
\lh \frac{\Lb^2}{\Lb_*^2} \rh^{\ag/2\pi},
\label{6.2}
\ee
where $g_*^2 = g_R^2(\Lb_*)$ at some reference scale $\Lb_*$. This 
solution has three branches, one for which $g_R^2 > \ag$, one for 
which $g_R^2 < - \ag$ and one for which $|g_R^2| < \ag$. 
The domains of these branches are determined by a scale $\Lb_s$ such 
that
\be 
\lh \frac{\Lb_s^2}{\Lb_*^2} \rh^{\ag/2\pi} = 
\left| \frac{g_*^2 + \ag}{g_*^2 - \ag} \right|.
\label{6.3}
\ee
Defining $g_s^2$ by  $g_s^2 \equiv g^2_R(\Lambda_s)$, it follows that 
\be
\left| \frac{g_s^2 + \ag}{g_s^2 - \ag} \right| = 1 \hs{1} \Rightarrow \hs{1}
g^2_s = 0, \pm \infty.
\label{6.4}
\ee
The various branches have been plotted in figures \ref{grplot2a} and 
\ref{grplot2b}. Figure \ref{grplot2a} shows both the branches for 
$g_R^2 > \ag$ and for $g_R^2 < - \ag$, having $g_s^2 = \pm\infty$. Figure \ref{grplot2b} shows the
branch for $|g_R^2| < \ag$, having $g_s^2 =0$. 

It is clear that in the limits $\Lambda \rightarrow 0$, 
$\Lambda \rightarrow \infty$ the renormalized self coupling $g_R^2$ 
approaches the two special constant values
\be
g_R^2(0) = \alpha, \quad g_R^2(\infty) = -\alpha.
\ee
In the limit $\alpha \rightarrow 0$, the branches with 
$g_R^2(\Lb_s) = \pm \infty$, and hence $|g_R^2(\Lb)| > \ag$, 
reduce to the solution found for the $\beta$-function of the non-linear 
Schr\"odinger model \cite{DeKokVanHolten}. This is as expected since the 
Jackiw-Pi model reduces to that model when taking $e_R = 0$ leading to 
$\alpha = 0$. Note, that for $|g_R^2(\Lambda_*)| > \alpha \geq 1$ the 
Jackiw-Pi model is never in the perturbative regime. However, when 
$\alpha < 1$ there is a perturbative domain $g_R^2(\Lb) < 1$ for both 
small and very large $\Lambda$.

The new branch with $g_R^2(\Lambda_s) = 0$ and $|g_R^2(\Lb)| < \ag$ has 
been plotted in figure \ref{grplot2b}.
From this solution it is clear that the model is in the perturbative 
regime for intermediate values of $\Lambda$ when $\alpha \geq 1$, 
and for all values of $\Lambda$ when $\alpha < 1$, at least in the 1-loop
approximation.

\section{Conclusion}
\label{JPconclusie}

It has been shown that to first order in perturbation theory only 
the coupling constant $g^2$ is renormalized: neither the fields, nor the 
electric charge are scale dependent. 
Moreover, if the coupling constants in the classical model are chosen 
such that $g^2 = \pm e^2$, this relation is preserved and $g^2 = g_R^2$ 
is constant as well. 

In all other cases the renormalized coupling constant 
$g_R^2$ becomes a function of the scale $\Lambda$ and the scale and 
special conformal symmetries are anomalous. Since only $g^2$ is 
renormalized, the argument given in \cite{DeKokVanHolten} makes it 
clear that the time-dependence of the corresponding charges $D$ and $K$ 
is given by 
\be
\frac{dK}{dt} & =& -t\frac{dD}{dt} ; \\\nonumber
\frac{dD}{dt} &=& \frac{1}{2}\, \beta(g_R^2)\, \int d^2x\, \Fg^{\dag\,2} \Fg^2 
 = \frac{1}{4\pi}\left(g_R^4-e_R^4\right) \int d^2x\, \Fg^{\dag\,2} \Fg^2.
\ee
From this expression it follows that the conformal symmetries indeed 
survive quantization in $1$-loop approximation if and only if the condition 
$g^2 = g_R^2 = \pm e_R^2$ is satisfied.

It so happens that for the minus sign this condition is exactly the one 
the coupling constants need to satisfy such that the classical self-dual 
vortex solutions exist (\ref{jw.1}). In this 
special case not only the conformal symmetries survive quantization, 
but also the classical self-dual solutions. The condition 
$g^2 = - e^2$ is thus essential for the existence of self-dual vortex 
solutions in the classical theory and in the quantum theory as well.

Assuming $\alpha = e^2$ to be less than unity, and choosing $g_R^2$ 
less than $\alpha$ at some reference energy scale $\Lambda_*$,
the Jackiw-Pi model is seen to have the unusual property that the model 
is in the perturbative regime for all values of $\Lambda$, as shown in 
figure  \ref{grplot2b}. For small values of $\Lambda$ $g_R^2$ is seen 
to be positive, signifying a repulsive interaction between the particles, 
and for large values of $\Lambda$ $g_R^2$ is negative, signifying an 
attractive interaction between the particles. 

In the case $\alpha$ less than unity, when $g_R^2$ is larger 
than $\alpha$ at the reference energy scale, the model has a perturbative 
regime either for small values of $\Lambda$ or for very large values, as 
shown in figure  \ref{grplot2a}. When $\alpha$ is larger than one, there 
only is a perturbative regime for intermediate values of $\Lambda$ when 
$g_R^2$ at the reference scale is chosen less than $\alpha$, as
shown in figure \ref{grplot2b}.


{\small\bibliography{PhDThesis}}

\end{document}